# Adaptive Decision-Objective Loss for Forecast-then-Optimize in Power Systems

**Anonymous Submission**

**Abstract**

Forecast-then-optimize is a widely-used framework for decision-making problems in power systems. Traditionally, statistical losses have been employed to train forecasting models, but recent research demonstrated that improved decision utility in downstream optimization tasks can be achieved by using decision loss as an alternative. However, the implementation of decision loss in power systems faces challenges in 1) accommodating multi-stage decision-making problems where upstream optimality cannot guarantee final optimality; 2) adapting to dynamic environments such as changing parameters and nature of the problem like continuous or discrete optimization tasks. To this end, this paper proposes a novel adaptive decision-objective loss (ADOL) to address the above challenges. Specifically, ADOL first redefines the decision loss as objective utilities rather than objective loss to eliminate the need to manually set the optimal decision, thus ensuring the globally optimal decision. ADOL enables one-off training in a dynamic environment by introducing additional variables. The differentiability and convexity of ADOL provide useful gradients for forecasting model training in conjunction with continuous and discrete optimization tasks. Experiments are conducted for both linear programming-based and mixed-integer linear programming-based power system two-stage dispatching cases with changing costs, and the results show that the proposed ADOL is capable of achieving globally optimal decision-making and adaptability to dynamic environments. The method can be extended to other multi-stage tasks in complex systems.

## Introduction

Forecast-then-optimize (Donti, Amos, and Kolter 2017) is a framework widely used in various decision-making problems under uncertainty. This framework operates in two distinct steps: forecasting which uses various features to provide forecasts for all uncertainties that affect the decision-making problem, and optimization which utilizes the forecast and obtains the optimal decision for a certain objective. Such tasks could be multi-stage including frequent updates of forecast and corrective decisions. Taking power system unit dispatching as an example, a power system is a network used to generate, transmit, and distribute electricity and unit commitment refers to the schedule of power generating units over a certain time frame to meet the electricity demand at the lowest possible cost. Electric loads are first forecasted day-ahead, and power units are then scheduled to meet the load demand plus some reserves, which are defined by rules such as a certain percentage of the total load demand (Ela et al. 2011). Finally, the reserves are utilized intra-day to compensate for unbalanced loads due to forecasting errors.

The two steps of forecast-then-optimize are traditionally decoupled where the forecasting aims to achieve the utmost statistical precision by minimizing the statistical loss and the optimization takes the outcome of forecasting as inputs for decision-making. An underlying assumption here is that statistically accurate forecasts will guarantee effective decision-making. However, recent research has demonstrated that the impact of forecasting errors on decision objectives could be asymmetric and non-linear (Carriere and Kariniotakis 2019; Muñoz, Pineda, and Morales 2022; Li and Chiang 2018), leading to a mismatch between more accurate forecasts and more beneficial decision-making.

Recent work in the forecast-then-optimize has shown the potential to align forecasting performance with downstream optimization tasks. This is achieved by replacing the traditional statistical loss with a decision loss, which measures the utility deviation between the decision based on the forecasted value and the decision based on the true value (Kotary et al. 2021). The approximation of the decision loss has been a central focus of previous research efforts (Ferber et al. 2009; Wilder, Dilkina, and Tambe 2019; Balghiti et al. 2022; Mulamba et al. 2020; Guler et al. 2022). The conventional method is to design task-specific surrogates for the original optimization task while more recent work has investigated data-driven methods to learn the decision loss directly (Zhang, Wang, and Hug 2022; Shah et al. 2022). However, the decision loss is designed with the assumption that decision-making is single-stage and static, posing hurdles to their application in multi-stage problems within complex dynamic systems such as power systems.

First, decision loss is unsuitable for multi-stage decision-making problems in complex systems where the initial forecasting error will propagate and be corrected along the stages. Corrective decisions exhibit varying tolerance levels in response to different forecasting errors, which may lead to locally optimal decisions, i.e., a perfect forecast may not lead to optimal decisions. For instance, power system operation requires initial day-ahead decisions (power plant output and reserve schedule) based on the day-ahead forecast and intra-day corrective decisions based on the updated forecast. Compared with an "error-free" perfect day-ahead forecast, the decision would be better off when the forecast slightly deviates and let the reserve cover the unbalanced loads at the intra-day stage (Chen, Liu, and Wu 2022). The fundamental reason is that the reserve is mandatory even under perfect forecast but may not be fully utilized in the intra-day operation. Second, decision loss is difficult to adapt to dynamic environments. In power systems, decision-making problems often require frequent updates within dynamic environments (characterized by unit cost changes) that feature constantly varying parameters. Within such environments, decision loss-based forecasting models require frequent retraining, which limits their practical applications. Third, continuous and discrete optimization tasks have different decision losses that may be non-convex and non-differentiable, which requires different approximation strategies. This may limit the wide applicability of the approach.

To tackle the challenges above, we propose a novel loss function, denoted as adaptive decision-objective loss (ADOL), for forecast-then-optimize tasks in power systems. Unlike conventional decision loss, ADOL relies on a neural network to directly learn the decision objective utilities under various forecasting errors without manually defining the optimal decision point, which was traditionally conceived as the decision under perfect forecast. To the best of the authors' knowledge, this is the first work that proposes to use a neural network as a loss function, with the following advantages: 1) Strong ability to adapt against dynamic environments by considering the impact of parameter variations on the mapping between forecasting errors and decision objectives. 2) Differentiability and convexity, which can provide useful gradients when facing continuous and discrete downstream optimization tasks. We verify the effectiveness of ADOL through two power system dispatching cases: one based on linear programming (LP) and the other based on mixed-integer linear programming (MILP).

## Problem Formulation and Related Work

In this section, we first illustrate how decision loss improves the decision utility of a single-stage decision-making problem, followed by the challenges with decision loss when applied to multi-stage decision-making problems (with an example of two-stage unit dispatch in power systems). Finally, we present the challenges of applying decision loss to dynamic environments and different optimization tasks.

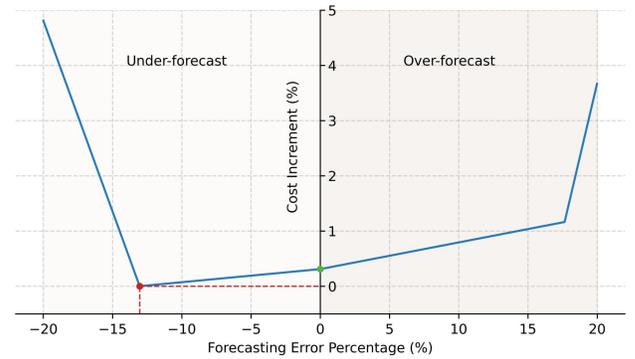

Figure 1: Cost increments for different forecasting error percentages in a two-stage unit dispatch case.

### Decision Loss for Single-Stage Decision-Making

Single-stage decision-making can be modeled as follows:

$$\mathcal{O}(\hat{y}, w) = \arg\min_{z} c(z; \hat{y}, w) \text{ s.t. } z \in \mathcal{C}(\hat{y}, w) \quad (1)$$

where $\hat{y}$ is the forecast of the true parameter vector $y$, $w$ is the known parameter vector, and $z$ is the decision vector. The decision objective $c(z; \hat{y}, w)$ and feasible region $\mathcal{C}(\hat{y}, w)$ depend on $\hat{y}$ and $w$. $\mathcal{O}(\hat{y}, w)$ represents the forecasted optimal decision vector given $\hat{y}$ and $w$. Traditionally, forecasting models for $y$ are trained by minimizing a statistical loss $\mathcal{L}(\hat{y}, y)$ (e.g. mean squared error) that is characterized by a minimum point at zero and symmetric about zero, which means over-forecast and under-forecast have the same loss. However, this may lead to sub-optimal decisions, since the impact of forecasting errors on the decision objective may be asymmetric and non-linear.

Decision loss $\mathcal{L}(\mathcal{O}(\hat{y}, w), \mathcal{O}(y, w))$ has been proposed to integrate the impact of forecasting errors on the decision objective in the process of forecasting model training, which represents the deviation between the forecasted optimal decision $\mathcal{O}(\hat{y}, w)$ and the target optimal decision $\mathcal{O}(y, w)$.

### Challenges with Decision Loss in Two-Stage Decision-Making

Unit dispatch in power systems (Dvorkin, Delikaraoglou, and Morales 2018) is usually modeled as a two-stage decision-making problem: In the day-ahead stage, generating units are scheduled to generate sufficient power to meet forecasted load demands, with certain upward and downward reserves to mitigate the under-forecast and over-forecast situations. In the intra-day stage, any deviation between the true load demand and the forecasted load demand is covered by proper balancing actions through activating the scheduled reserves in the day-ahead stage. These two stages usually take the system operating cost as the decision objective, as follows:

Day-ahead stage:
$$\mathcal{O}_1(\hat{y}, w_1) = \arg\min_{z_1} c_1(z_1; \hat{y}, w_1) \text{ s.t. } z_1 \in \mathcal{C}(\hat{y}, w_1) \quad (2.1)$$

Intra-day stage:
$$\min_{z_2} c_2(z_2; y, w_2, \mathcal{O}_1(\hat{y}, w_1)) \text{ s.t. } z_2 \in \mathcal{C}(y, w_2, \mathcal{O}_1(\hat{y}, w_1)) \quad (2.2)$$

In the above decision-making process, the total cost $c_1 + c_2$ depends on the forecasting error ($\hat{y} - y$). Interestingly, achieving a perfect forecast ($\hat{y} = y$) does not necessarily lead to the optimal total cost in some scenarios. We illustrate this with a specific case (see Appendix A), where the relationship between forecasting error percentage and cost increment is shown in Figure 1. It can be seen that the cost increment incurred by a perfect forecast (green point) is higher than that incurred by an under-forecast (red point). In this case, training a forecasting model by minimizing the decision loss leads to locally optimal decisions, because minimizing the decision loss is to minimize the deviation between the forecasted optimal solution and the target optimal solution, rather than to minimize the decision objective.

### Challenges with Decision Loss in Dynamic Environments

In certain power system applications, the parameter vector $w$ will change dynamically such as generation cost. In this dynamic environment, decision loss-based forecasting models face the challenge of necessitating frequent retraining, which requires solving one optimization task for each training instance at every epoch. This significantly increases the computational burden, resulting in the implementation of decision loss in power system applications, typically constrained by time, a difficult task. Although Shah et al. (Shah et al. 2022) and Zhang et al. (Zhang, Wang, and Hug 2022) use sampled data to learn the data-driven decision loss for optimization-free forecasting model training, in dynamic environments, such data-driven decision loss still requires frequent relearning, which may lead to a higher computational burden due to the data sampling intricacies.

### Challenges with Decision Loss in Facing Continuous and Discrete Optimization Tasks

The non-convex and non-differentiable properties of decision loss result in gradients often being zero or undefined, which is unsuitable for training forecasting models. The existing literature proposes different approximation strategies for continuous and discrete optimization tasks to provide useful gradients (Kao, Roy, and Yan 2009; Wilder, Dilkina, and Tambe 2019; Balghiti et al. 2022; Mulamba et al. 2020; Guler et al. 2022; Mandi and Guns 2020; Elmachtoub and Grigas 2021; Mandi et al. 2020). These strategies include relaxing the original problem, introducing regularization, etc. However, the need to set different approximation strategies for each optimization task limits the applicability of decision loss, as utilities typically manage multiple decision-making problems, which may be based on LP, MILP, etc.

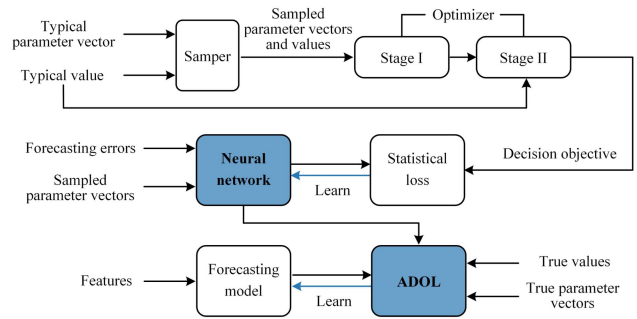

Figure 2: The overall process from learning ADOL to training the forecasting model.

## Learning ADOL for Forecasting Model Training

The idea behind ADOL is twofold: 1) redefining the decision loss as objective utilities rather than objective loss to eliminate the need for manual setting of the optimal decision, and 2) utilizing the neural network to replace the linear loss function framework (Shah et al. 2022; Zhang, Wang, and Hug 2022). This enables one-off training by introducing additional variables of the dynamic environments, as well as providing useful gradients for continuous and discrete optimization tasks based on the convexity of neural networks and the ability to fit non-linear mappings.

Specifically, ADOL employs a neural network to fit the mapping relationship between the forecasting error and the total decision objective, which is then used as a loss function to guide the forecasting model training. In this way, training the forecasting model is to minimize the total decision objective as the ultimate goal, thereby avoiding locally optimal decisions. Although training a neural network is not a convex problem, the trained neural network itself is convex as a loss function. This can be proved by the positive semi-definiteness of the Hessian matrix of the neural network (Bertsekas, D. 2009), thus providing useful gradients for forecasting model training when faced with continuous and discrete optimization tasks. Moreover, due to the scalability of neural networks, ADOL can achieve one-off learning by considering the impact of the parameter vector $w = (w_1, w_2)$ on the mapping relationship to adapt against dynamic environments. The overall process from learning the ADOL to training the forecasting model is shown in Figure 2.

**Sampling:** First, the typical value $\bar{y}$ and the typical parameter vector $\bar{w}$ are generated from the historical dataset $[y_1,...,y_N]$ and $[w_1,...,w_N]$, respectively. Secondly, a set of $K$ points $[\hat{y}_1,...,\hat{y}_K]$ and $[\hat{w}_1,...,\hat{w}_K]$ are sampled around $\bar{y}$ and $\bar{w}$, respectively. $\hat{y}_k$ and $\hat{w}_k$ are randomly selected in the interval controlled by $\gamma$ and $\beta$, respectively:

$$\hat{y}_k \in [(1-\gamma)\bar{y}, (1+\gamma)\bar{y}] \quad (3.1)$$
$$\hat{w}_k \in [(1-\beta)\bar{w}, (1+\beta)\bar{w}] \quad (3.2)$$

The generation method for $\bar{y}$ and $\bar{w}$, as well as the selection of $\gamma$ and $\beta$, can be arbitrary, but it is essential to attempt to cover all possible scenarios. For instance, for load forecasting, the typical value $\bar{y}$ can be a representative daily load profile. $\gamma$ can take a value less than 0.2, because the deviation of the load forecasting usually does not exceed ±20%. Second, we calculate the total decision objective for each sample instance. Specifically, we input $\bar{y}$, $\hat{y}$, and $\hat{w} = (\hat{w}_1, \hat{w}_2)$ into the optimizer, and obtain the total decision objective $[c^1,...,c^K]$ by solving the two-stage optimization model (4.1)-(4.2), where $c = c_1 + c_2$.

Stage I:
$$\mathcal{O}_1(\hat{y},\hat{w}_1) = \arg\min_{z_1} c_1(z_1; \hat{y}, \hat{w}_1) \text{ s.t. } z_1 \in \mathcal{C}(\hat{y},\hat{w}_1) \quad (4.1)$$

Stage II:
$$\min_{z_2} c_2(z_2; \bar{y}, \hat{w}_2, \mathcal{O}_1(\hat{y},\hat{w}_1)) \text{ s.t. } z_2 \in \mathcal{C}(\bar{y}, \hat{w}_2, \mathcal{O}_1(\hat{y},\hat{w}_1)) \quad (4.2)$$

**Learning ADOL:** The neural network input is the forecasting error $\hat{y} - \bar{y}$ and the parameter vector $\hat{w}$, and the output is the corresponding total decision objective $c$. The trained neural network can be regarded as a function of $\hat{y} - \bar{y}$ and $\hat{w}$, defended as $\hat{c}(\hat{y} - \bar{y}, \hat{w}; \omega)$, where $\omega$ represents the parameters in the neural network, obtained by minimizing the statistical loss such as mean square error (MSE):

$$\omega^{i+1} = \omega^i - \alpha \frac{\partial (\hat{c}(\hat{y} - \bar{y}, \hat{w}; \omega) - c)^2}{\partial \omega^i} \quad (5)$$

where $\alpha$ is the learning rate and $i$ is the number of iterations. It is worth noting that the learning of ADOL is one-off. When the dynamic parameters change, it is only necessary to adjust the parameter vector $w$ in ADOL and then retrain the forecasting model.

**Training forecasting models:** Given the parameter vector $w$, the update process of the parameter $\theta$ of the forecasting model $f(x,\theta)$ is as follows:

$$\theta^{i+1} = \theta^i - \alpha \frac{\partial \hat{c}(f(x,\theta^i) - y, w; \omega)}{\partial \theta^i} \quad (6)$$

# Experiments

## Case Description

Here, we provide a conventional unit dispatch case for power systems. In this case, day-ahead load forecasting will be performed first, then unit dispatching will be performed in the day-ahead electricity market to meet the load demand, and finally unit re-dispatching will be performed in the real-time electricity market to cover the unbalanced load, i.e., the deviation between the forecasted load and the true load. All cases are solved on a PC with the Intel i5-12400F CPU @ 2.50 GHz, NVIDIA GTX 1660Ti 6GB, and 16GB RAM.

**Forecast (load forecasting):** We verify the performance of the proposed method on the Panama region load forecasting dataset provided by Aguilar Madrid et al. (Aguilar Madrid, E.; and Antonio, N. 2021) The dataset includes various features such as historical electricity loads, weather information for three main cities in Panama, calendar information related to holidays and semesters, etc. The time granularity of these data is 1 hour. Since feature selection is not the main focus of this paper, we do not perform feature selection, only normalization. In this paper, the training set is based on the data of 2015, the forecasting horizon is from 2016/1/1 to 2016/1/31 with a resolution of 1 hour.

**Optimize (electricity market clearing):** The specific electricity market clearing model is shown in Case 1 in the Appendix. the specific parameters are set as follows: $\mathcal{I} = 3$, $\mathcal{J} = 3$, $C = [30, 40, 50]$ (\$), $C^{qs} = [60, 70, 80]$ (\$), $C^U = C \times 10\%$, $C^D = C \times 2\%$, $S^{U/D} = 0.15$, $\bar{P} = [800, 800, 800]$ (MW), $\bar{R}^{U/D} = \bar{P} \times 40\%$, $\bar{P}^{qs} = [200, 200, 200]$ (MW).

## Evaluation Criteria

Two different evaluation criteria are used to evaluate the performance of the proposed ADOL. The first is used to evaluate the economic efficiencies caused by forecasts, and the last one is used to evaluate the statistical errors of forecasts.

**Mean Total Decision Objective (MTDO):** The MTDO is the mean total decision objective for all testing instances, calculated by solving (2.1) and (2.2), as below:

$$\frac{1}{T}\sum_t^T (c_1^t + c_2^t) \quad (7)$$

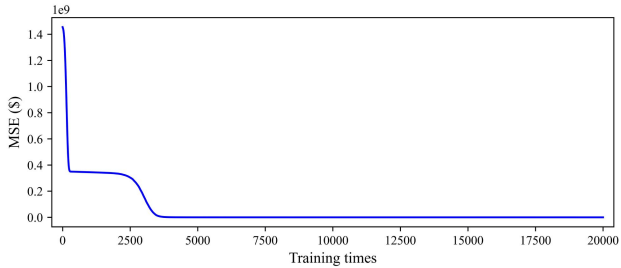

Figure 3: The change of MSE during ADOL learning.

| Loss Function | MAPE (%) (mean/std) | MTDO ($) (mean/std) |
|---|---|---|
| Optimal forecast | 13.04 | **38369.92** |
| Perfect forecast | **0** | 38487.56 |
| ADOL | 10.58 (0.41) | **38712.43 (41.17)** |
| DL | 10.21 (0.37) | 38928.91 (39.12) |
| MSE | **9.98 (0.21)** | 39539.94 (32.31) |

Table 1: Comparative analysis with different forecasting.

Where $T$ represents the number of instances in the test set and $c_1^t + c_2^t$ represents the total decision objective caused by the test instance $t$. For minimization problems, the smaller MTDO is better, while for maximization problems, the larger MTDO is better.

**Mean Absolute Percentage Error (MAPE):** The MAPE is used to evaluate the error percentage between the forecasted value and the true value, which can be calculated as below:

$$\frac{1}{T}\sum_{t}^{T}(\frac{\hat{y}_t - y_t}{y_t} \times 100\%) \qquad (8)$$

## ADOL for LP-Based Cases

**ADOL training:** We first generated 20,000 sample points based on (3.1) using the average value of the 2015 electricity load in the Panama region as the typical load $\bar{L}$, where $\gamma = 0.2$. These samples, as well as the typical load, were then fed into the day-ahead electricity market model (9) as well as the real-time electricity market model (10) to calculate the total cost. Finally, a 2-layer fully connected neural network with 100 hidden units was trained with load forecast error $\bar{L} - L$ as input and total cost as output to obtain ADOL. The training process of ADOL is shown in Figure 3. With the increase in training times, the MSE decreases. After 20,000 times of training, the MSE is stable at around 230$, and the MAPE is stable at around 0.02%.

**Comparison with other forecasts:** Three forecasting models are used to forecast the electric load from 2016/1/1 to 2016/1/31. The three forecasting models are all based on a 2-layer fully connected neural network with 100 hidden units, the first model is trained based on ADOL (ADOL), the second model is trained based on decision loss (DL) (Chen et al. 2021), and the third model is trained based on mean square error loss (MSE). We run each model 10 times, and we report the mean and standard deviation (in brackets) of the 10 runs. In addition, a comparison is made with two theoretically optimal forecasts: a perfect forecast in which the forecast is equal to the true value (Perfect), and an objective-optimal forecast is one in which the forecast makes the decision objective optimal (Optimal).

The main results are shown in Table 1.

In this case, the optimal forecast exhibits better decision utility, with a 0.31% reduction in MTDO, although the statistical error of the optimal forecast is 13.04% higher than that of the perfect forecast. The ADOL-based forecasting model outperforms the DL-based forecasting model (0.56% reduction in MTDO) and the MSE loss-based forecasting model (2.09% reduction in MTDO) in terms of decision utility, because its ultimate goal is to achieve optimal forecasting rather than perfect forecasting.

**Forecasting error analysis:** We conducted an error analysis on three forecasting models: the ADOL-based forecasting model, the DL-based forecasting model, and the MSE loss-based forecasting model. The results are illustrated in Figure 4. In Figure 4(a), we present the mapping relationship between forecasting error and cost increment in this case. The red point in the figure represents the optimal forecast point with -13.04% forecasting error and the lowest cost increment. At this point, the capacity of the upward reserve is 13.04% of the true value of the load, which can just cover the unbalanced load of 13.04%. Figures 4(b), 4(c), and 4(d) display the forecasting error distributions of the ADOL-based, DL-based, and MSE loss-based forecasting models on the test set, respectively, represented by histograms. Upon observation, we noticed that the ADOL-based forecasting errors are primarily concentrated near the red point, which indicates that our method can effectively achieve near-optimal decisions. DL-based forecasting errors tend to exhibit a negative error due to the smaller utility deviation when facing negative errors compared to positive errors. However, DL-based forecasting errors do not concentrate around the red point, leading to locally optimal decisions. Lastly, the MSE loss-based forecasting errors tend to be concentrated around the origin, reflecting their tendency to produce errors close to zero.

## Analysis of Adaptability of ADOL in Dynamic Environment

We consider the change in the parameter vector $w$ when training ADOL. In this case, $w$ consists of $C$, $C^{qs}$, $C^U$, and $C^D$. $\beta$ is set to 0.1, which means that the parameter varies within a range of ±10%. When testing the performance of forecasting models, 5 scenarios are generated by random sampling in the range of $[0.9w, 1.1w]$ for comparing the training time and MTDO of the ADOL-based forecasting model, the DL-based forecasting model, and the MSE loss-based forecasting model. All forecasting models are trained for 20,000 iterations, and the main results are shown in Table 2.

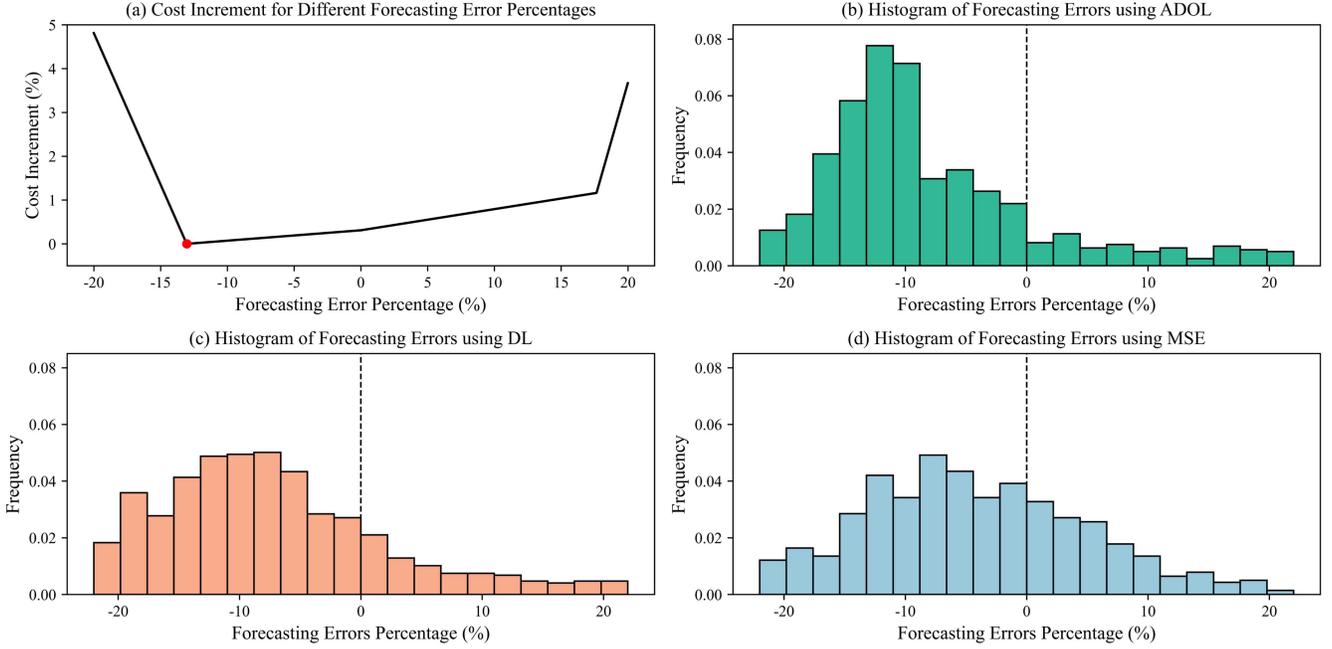

Figure 4: Error distribution for forecasting methods with different loss functions.

| Scenario | ADOL (mean/std) | DL (mean/std) | MSE (mean/std) |
|---|---|---|---|
| 1 | **40156.52 (31.81)** | 40185.12 (22.25) | 40320.76 (27.14) |
| 2 | **38783.28 (39.52)** | 39000.36 (48.01) | 39181.69 (37.79) |
| 3 | **40184.53 (43.34)** | 40262.22 (37.51) | 40665.83 (23.18) |
| 4 | **36625.14 (40.01)** | 36632.34 (30.16) | 37039.79 (23.47) |
| 5 | **41110.33 (45.68)** | 41280.46 (43.62) | 41378.29 (34.80) |
| Training time (s) | 27.08 (2.31) | 46.30 (2.21) | **24.71 (2.45)** |

Table 2: Comparative analysis on MTDO ($) and training time (s) in 5 scenarios.

|  | ADOL (mean/std) | DL (mean/std) | MSE (mean/std) |
|---|---|---|---|
| MTDO ($) | **397883.92 (597.61)** | 399077.08 (619.03) | 401849.67 (534.16) |
| Training time (s) | 29.36 (1.13) | 25420 (96) | **25.41 (0.98)** |

Table 3: Comparative analysis for a MILP-based case.

ADOL consistently performs well on MTDO. Compared with DL, ADOL achieved an average reduction of 0.25% on MTDO. Similarly, compared with MSE loss, ADOL also achieved an average reduction of 0.87% on MTDO. As for training time, the time required to train the forecasting model based on ADOL is comparable to that of MSE loss, with ADOL being only 9.59% higher than MSE loss. However, the training time for the forecasting model based on ADOL is significantly lower than that for DL, with ADOL requiring only 58.49% of the training time of DL. This computational advantage becomes more pronounced when dealing with more complex cases, as demonstrated in the following Subsection.

### ADOL for MILP-Based Cases

A MILP-based unit dispatch example is utilized to evaluate ADOL, as presented in Case 2 in the Appendix B. Compared to Case 1, Case 2 incorporates considerations for unit startup and shutdown costs, as well as the correlation between time series. The case is conducted on an IEEE 30-node system (Reddy and Rathnam 2016), and the Panama data is properly scaled to generate load data for each node in the system. In this case, we compare ADOL-based, DL-based, and MSE loss-based forecasting models. All forecasting models are trained for 20,000 iterations, and the main results are shown in Table 3.

In this case, ADOL achieves a 0.3% reduction in MTDO compared to DL and a 0.99% reduction in MTDO compared to MSE loss. In terms of training time, ADOL-based forecasting model training time is still comparable to MSE loss, at 29.36s and 25.41s, respectively. However, the training time of DL-based forecasting models is 865 times higher than that of ADOL-based forecasting model, as the optimization model needs to be solved in each iteration. Although Zhang et al. and Li et al. proposed optimization-free DL to significantly speed up forecasting model training, DL still requires relearning when the environment changes. In contrast, since ADOL considers the impact of environment changes, its training is one-off.

## Conclusion

In this paper, a neural network-based loss function, ADOL, is proposed for multi-stage tasks in complex systems. It simultaneously address the three key challenges of 1) ensuring globally optimal decisions, 2) adapting to dynamic environments, and 3) being applicable to continuous and discrete optimization tasks. The proposed ADOL is verified in two power system dispatching cases based on LP and MILP as downstream decision-making problems. The results show that ADOL-based forecasts can further improve decision utility compared with DL-based forecasts. In addition, the training time required to train an ADOL-based forecasting model comparable to that required to train a MSE loss-based forecasting model, and significantly lower than that required to train a DL-based forecasting model. This benefit of ADOL becomes more significant when the downstream optimization model becomes more complex. In future work, the proposed ADOL will be applied to more general multi-stage decision-making problems, beyond power system applications.

## Appendix A: Case 1

This is a unit dispatching model based on linear programming (Viafora et al. 2020). Given the forecasted value of the net load $\hat{L}$, the lowest-cost day-ahead energy schedule can be formulated as follows:

$$\min_{P_i, R_i^U, R_i^D} \sum_i^{\mathcal{I}} (C_i P_i + C_i^U R_i^U + C_i^D R_i^D) \quad (9.1)$$

$$\text{s.t.} \sum_i^{\mathcal{I}} P_i = \hat{L} \quad (9.2)$$

$$\sum_i^{\mathcal{I}} R_i^U = S^U \hat{L}, \quad \sum_i^{\mathcal{I}} R_i^D = S^D \hat{L} \quad (9.3)$$

$$R_i^U + R_i^D \leq \overline{P}_i, \quad \forall i \in \mathcal{I} \quad (9.4)$$

$$0 \leq R_i^U \leq \overline{R}_i^U, \quad 0 \leq R_i^D \leq \overline{R}_i^D, \quad \forall i \in \mathcal{I} \quad (9.5)$$

$$R_i^D \leq P_i \leq \overline{P}_i - R_i^U, \quad \forall i \in \mathcal{I} \quad (9.6)$$

where $\mathcal{I}$ represents the number of units. $P_i$ and $C_i$ represent the power generation and cost coefficients of unit $i$, respectively. $R_i^{U/D}$ represents the up-/downward reserve provision of unit $i$, while $C_i^{U/D}$ represents the corresponding up-/downward reserve cost coefficients of unit $i$. $\overline{P}_i$ represents the power generation limit of unit $i$. $\overline{R}_i^{U/D}$ represents the up-/downward reserve limit of unit $i$. Reserve requirements are based on a certain percentage of the net load, with the proportional coefficients of up-/downward reserves being represented as $S^{U/D}$. The decision objective (9.1) is to minimize day-ahead costs. Constraint (9.2) ensures the balance of the load demand. Constraints (9.3) ensure that upward and downward reserve requirements are fulfilled, whereas constraints (9.4)-(9.5) consider the upper limit of reserve provided by each unit. Constraints (9.6) consider the generation limits of each unit. Once the true load demand $L$ is realized, and the reserve allocation $R^{U*}$ and $R^{D*}$ are determined, the lowest-cost real-time energy schedule can be formulated as:

$$\min_{P_j^{qs}, r_i^U, r_i^D} \sum_i^{\mathcal{I}} C_i (r_i^U - r_i^D) + \sum_j^{\mathcal{J}} C_j^{qs} P_j^{qs} \quad (10.1)$$

$$\text{s.t.} \sum_j^{\mathcal{J}} P_j^{qs} + \sum_i^{\mathcal{I}} (r_i^U - r_i^D) \geq L - \hat{L} \quad (10.2)$$

$$0 \leq P_j \leq \overline{P}_j^{qs}, \quad \forall j \in \mathcal{J} \quad (10.3)$$

$$0 \leq r_i^U \leq R_i^{U*}, \quad 0 \leq r_i^D \leq R_i^{D*}, \quad \forall i \in \mathcal{I} \quad (10.4)$$

where $\mathcal{J}$ represents the number of quick-start units. $P_j^{qs}$ and $C_j^{qs}$ represent the power generation and cost coefficients of quick-start unit $j$, respectively. $r_j^{U/D}$ represents the up-/downward reserve deployment of unit $j$. $\overline{P}_j^{qs}$ represents the power generation limit of quick-start unit $j$. The decision objective (10.1) is to minimize real-time costs. Constraint (10.2) ensures the real-time power balance. Constraints (10.3) consider the generation limits of each quick-start unit. Constraints (10.4) limit the activation of upward and downward reserves considering the procured reserve quantities.

## Appendix B: Case 2

This is a unit dispatching model based on mixed-integer linear programming (Chen, Liu, and Wu 2022). The

lowest-cost day-ahead energy schedule can be formulated as follows:

$$\min_{\Xi} \sum_{t}^{\mathcal{T}} \sum_{i}^{\mathcal{I}} (C_i P_{it} + C_i^{\text{U}} R_{it}^{\text{U}} + C_i^{\text{D}} R_{it}^{\text{D}} + C_i^{\text{SU}} U_{it} + C_i^{\text{SD}} D_{it}) \quad (11.1)$$

s.t. $\sum_{i}^{\mathcal{I}} P_{it} - \sum_{k}^{\mathcal{K}} \hat{L}_{kt} - \sum_{m:(n,m)\in\Lambda} \dfrac{\delta_{nt}^{\text{DA}} - \delta_{mt}^{\text{DA}}}{x_{nm}} = 0, \quad \forall n \in \mathcal{N}, \forall t \in \mathcal{T}$ (11.2)

$\dfrac{\delta_{nt}^{\text{DA}} - \delta_{mt}^{\text{DA}}}{x_{nm}} \leq \overline{F}_{nm}, \quad \forall(n,m)\in\Lambda, \forall t \in \mathcal{T}$ (11.3)

$\sum_{i}^{\mathcal{I}} R_{it}^{\text{U}} = S^{\text{U}} \sum_{k}^{\mathcal{K}} \hat{L}_{kt}, \quad \sum_{i}^{\mathcal{I}} R_{it}^{\text{D}} = S^{\text{D}} \sum_{k}^{\mathcal{K}} \hat{L}_{kt}, \quad \forall t \in \mathcal{T}$ (11.4)

$P_{it} + R_{it}^{\text{U}} \leq P_i^{\max} I_{it}, \quad \forall i \in \mathcal{I}, \forall t \in \mathcal{T}$ (11.5)

$P_{it} - R_{it}^{\text{D}} \geq P_i^{\min} I_{it}, \quad \forall i \in \mathcal{I}, \forall t \in \mathcal{T}$ (11.6)

$U_{it} - D_{it} = I_{it} - I_{it-1}, \quad \forall i \in \mathcal{I}, \forall t \in \mathcal{T}$ (11.7)

$\sum_{t'=t-T_i^{su}+1}^{t} U_{it'} \leq I_{it}, \quad \forall i \in \mathcal{I}, \forall t \in \mathcal{T}_i^{su}$ (11.8)

$\sum_{t'=t-T_i^{sd}+1}^{t} D_{it'} \leq 1 - I_{it}, \quad \forall i \in \mathcal{I}, \forall t \in \mathcal{T}_i^{sd}$ (11.9)

$P_{it} - P_{i,t-1} \leq UR_i, \quad \forall i \in \mathcal{I}, \forall t \in \mathcal{T}$ (11.10)

$P_{i,t-1} - P_{it} \leq DR_i, \quad \forall i \in \mathcal{I}, \forall t \in \mathcal{T}$ (11.11)

$I_{it}, U_{it}, D_{it} \in \{0,1\} \quad \forall i \in \mathcal{I}, \forall t \in \mathcal{T}$ (11.12)

where $\mathcal{T}$ represents the length of time, $\mathcal{K}$ represents the number of nodes in the power system. $C_i^{\text{SU/SD}}$ represents the startup/shutdown cost coefficients of unit $i$. $P_i^{\min/\max}$ represents the minimum/maximum generation of unit $i$. $\delta_{nt}^{\text{DA}}$ represents the day-ahead voltage angle at node $n$ at time $t$. $\overline{F}_{nm}$ represents the capacity of transmission line $(n, m)$. $\mathcal{T}_i^{su/sd}$ is defined as $\{T_i^{su},...,T\}/\{T_i^{sd},...,T\}$, where $T_i^{su/sd}$ represents the minimum on/off time requirement of unit $i$. $I_{it}$ represents the unit commitment status of unit $i$ at hour $t$. $U_{it}/D_{it}$ represents the shutdown/startup status of unit $i$ at hour $t$. $UR_i/DR_i$ represents the startup/shutdown ramping capacity of unit $i$. The objective (11.1) is to minimize the total operating costs including generation, startup, shutdown, upward reserve, and downward reserve costs. System constraints include power balance (11.2), transmission line capacity constraints (11.3), and system reserve requirements (11.4). Unit constraints include generation limits (11.5)-(11.6), startup-shutdown status logic (11.7), minimum on and off requirements (11.8)-(11.9), ramping limits (11.10)-(11.11), and integrality requirements (11.12). Once the variables $\delta^{\text{DA}*}$, $R^{U*}$, and $R^{D*}$ are determined, the lowest-cost real-time energy schedule can be formulated as:

$$\min_{\Xi} \sum_{t}^{\mathcal{T}} \sum_{i}^{\mathcal{I}} C_i(r_{it}^{\text{U}} - r_{it}^{\text{D}}) + \sum_{t}^{\mathcal{T}} \sum_{j}^{\mathcal{J}} (C_j^{qs} P_{jt}^{qs} + C_j^{\text{SU-}qs} U_{jt}^{qs} + C_j^{\text{SD-}qs} D_{jt}^{qs})$$

(12.1)

s.t. $\sum_{i}^{\mathcal{I}} (r_{it}^{\text{U}} - r_{it}^{\text{D}}) + \sum_{k}^{\mathcal{K}} (\hat{L}_{kt} - L_{kt})$

$- \sum_{m:(n,m)\in\Lambda} \dfrac{\delta_{nt}^{\text{RT}} - \delta_{nt}^{\text{DA}*} - \delta_{mt}^{\text{RT}} + \delta_{mt}^{\text{DA}*}}{x_{nm}} = 0, \quad \forall n \in \mathcal{N}, \forall t \in \mathcal{T}$

(12.2)

$\dfrac{\delta_{nt}^{\text{RT}} - \delta_{mt}^{\text{RT}}}{x_{nm}} \leq \overline{F}_{nm}, \quad \forall(n,m)\in\Lambda, \forall t \in \mathcal{T}$ (12.3)

$0 \leq r_{it}^{\text{U}} \leq R_{it}^{\text{U}*}, \quad 0 \leq r_{it}^{\text{D}} \leq R_{it}^{\text{D}*}, \quad \forall i \in \mathcal{I}, \forall t \in \mathcal{T}$ (12.4)

$P_j^{\min,qs} I_{jt}^{qs} \leq P_{jt}^{qs} \leq P_j^{\max,qs} I_{jt}^{qs}, \quad \forall j \in \mathcal{I}, \forall t \in \mathcal{T}$ (12.5)

$U_{jt}^{qs} - D_{jt}^{qs} = I_{jt}^{qs} - I_{j,t-1}^{qs}, \quad \forall j \in \mathcal{I}, \forall t \in \mathcal{T}$ (12.6)

$P_{jt}^{qs} - P_{j,t-1}^{qs} \leq UR_j^{qs}, \quad \forall j \in \mathcal{I}, \forall t \in \mathcal{T}$ (12.7)

$P_{jt}^{qs} - P_{j,t-1}^{qs} \leq DR_j^{qs}, \quad \forall j \in \mathcal{I}, \forall t \in \mathcal{T}$ (12.8)

where $qs$ represents the quick-start unit, $\delta_{nt}^{\text{RT}}$ represents the real-time voltage angle at node $n$ at time $t$. The objective (12.1) is to minimize the total operating cost including the generation costs of all units, and the startup and shutdown costs of quick-start units. Constraint (12.2) ensures real-time nodal power balance. Constraints (12.3) provide the transmission line capacity limits. Constraints (12.4) limit the activation of upward and downward reserves. Constraints (12.5)-(12.8) are constraints for the quick-start unit.